# Shell Effects in Superdeformed Minima

P.-H. Heenen,[1,2] J. Dobaczewski,[2−4] W. Nazarewicz,[3−5] P. Bonche,[6] T.L. Khoo,[7]

[1]*Service de Physique Nucléaire Théorique, U.L.B - C.P. 229, B-1050 Brussels, Belgium*

[2]*Joint Institute for Heavy Ion Research, Oak Ridge National Laboratory, P.O. Box 2008, Oak Ridge, TN 37831, U.S.A.*

[3]*Department of Physics, University of Tennessee, Knoxville, TN 37996, U.S.A.*

[4]*Institute of Theoretical Physics, Warsaw University, Hoża 69, PL-00681, Warsaw, Poland*

[5]*Physics Division, Oak Ridge National Laboratory, P.O. Box 2008, Oak Ridge, TN 37831, U.S.A.*

[6]*Service de Physique Théorique, Centre d'Etudes de Saclay, 91191 Gif sur Yvette Cedex, France*

[7]*Physics Division, Argonne National Laboratory, Argonne, IL 60439, U.S.A.*

## Abstract

Recent experimental observation of the direct links between superdeformed and normal-deformed structures in the $A{\sim}190$ mass region offers a unique information on the absolute nuclear binding energy in the 2:1 minima, and hence on the magnitude of shell effects in the superdeformed well. In the present paper, the self-consistent mean-field theory with density-dependent pairing interaction is used to explain at the same time the two-particle separation energies in the first and second wells, and the excitation energies of superdeformed states in the $A{\sim}190$ and $A{\sim}240$ mass regions.

PACS numbers: 21.10.Ky, 21.10.Re, 21.60.Cs, 21.60.Jz

Typeset using REVTEX

# I. INTRODUCTION

Recent progress in gamma-ray spectroscopy with large gamma-ray detector arrays has resulted in the discovery of discrete lines linking superdeformed (SD) bands to low-deformation states. Transitions have been found in $^{194}$Hg [1] and $^{194}$Pb [2], which connect SD and normal-deformed states in one step, allowing the excitation energies, spins and likely parities of SD states to be determined. These quantities are only tentatively known in $^{192}$Hg, where high-energy transitions have been observed but have not been placed in the decay scheme of the SD band [3], and in $^{192}$Pb where one connecting transition has been tentatively assigned [4].

An important implication of these measurements is that it has became possible, for the first time, to establish experimentally the two-neutron and two-proton separation energies in SD minima. This study aims at analyzing these experimental results and at foreseeing the impact similar discoveries may have in the future on our understanding of nuclear shell properties.

The role of shell effects is well recognized in nuclear structure physics [5,6]. A decreased density of single-particle states around the Fermi level always leads to an increased stability of nuclear systems, in close analogy to phenomena known from atomic and molecular physics. The shell effects are, therefore, intimately related to the mean-field approximation, to which the very notion of individual particle orbits is inherent.

The observation of SD states constitutes an important confirmation of the shell structure of the nucleus. Quantum-mechanically, the remarkable stability of SD states can be attributed to strong shell effects that are present in the average nuclear potential at very elongated shapes [5,7–9]. For the oscillator potential this happens when the frequency ratio is 2:1 (for more realistic average potentials strong shell effects appear even at lower deformations). The structure of single-particle states around the Fermi level in SD nuclei is significantly different from the pattern at normal deformations. Indeed, the SD shells consist of states originating from spherical shells having different principal quantum numbers, hence having very different spatial character. Another interesting expected feature of the single-particle SD spectrum is a beating pattern in the level density, and hence in the shell-correction energy, giving rise to the so-called "super-shell" structure [6,5]. For nuclear ground-state configurations, the predicted period of beating is very long, hence impossible to see, considering the rather limited range of particle numbers available experimentally. On the other hand, it is believed that the beating pattern in SD states is particularly simple and its period is short, with the super-shell consisting of two neighboring SD shells only. Super-shell structure has been observed in metal clusters [10] where large electron numbers are accessible experimentally; it is consistent with the analysis based on a one-body finite potential [11]. The systematic measurements of binding energies of SD states will certainly shed some light on the super-shell structure of the deformed average field.

The main aim of this paper is to investigate the ability of the self-consistent mean-field approaches with realistic effective forces to explain *at the same time*: (i) the ground-state particle separation energies, (ii) the particle separation energies in SD minima, and (iii) the excitation energies of SD states in nuclei around $^{194}$Hg and $^{238}$U. The theoretical analysis is based on the self-consistent Hartree-Fock-Bogolyubov (HFB) approaches with effective Skyrme interactions. The details of our calculations are given in Sec. II, the results are presented in Sec. III, and Sec. IV contains summary and conclusions.

## II. METHOD OF CALCULATION

The calculations for separation energies in semi-magic nuclei (presented in Sec. III A) have been carried out within the spherical HFB approach of Ref. [12] with two Skyrme forces, SkP [12] and SLy4 [13,14]. In the latter case, we use in the pairing channel the density-dependent zero-range interaction

$$V_P = \frac{V_0}{2}(1 - P_\sigma)\left[1 - \frac{\rho(\boldsymbol{r}_1)}{\rho_c}\right]\delta(\boldsymbol{r}_1 - \boldsymbol{r}_2) \quad (1)$$

with parameters $\rho_c$ and $V_0$ adjusted within the method presented in Ref. [15]. In Eq. (1), $\rho(\boldsymbol{r})$ is the total local single-particle density in coordinate space.

The deformed calculations (Sec. III B-III D) have been performed with the HFB+LN method presented in Ref. [16]. Three Skyrme forces, namely SkP, SLy4, and SkM* [17] have been employed in the particle-hole channel. The SkM* force has been specifically adjusted to the fission barrier of $^{240}$Pu, and it has been proved in numerous studies of deformation effects to be quite successful. Recently, the SkM* force has been employed to describe the SD minima in the Hg-Pb region [18] and in the actinides [19]. Predictions of the more recent forces SkP and SLy4 for deformation effects have not yet been studied extensively.

The density-dependent pairing interaction of Eq. (1) has been used with a strength $V_0$ modified as compared to spherical HFB calculations because of the inclusion of different pairing spaces. It has been shown in Ref. [20] that changes in the size of the pairing space lead to uncertainty in the total energies of the order of a few hundred keV. For SkM*, the value $V_0 = -880$ MeV fm$^{-3}$ has been taken (see Ref. [16]) with a cut-off in the active pairing space of 5 MeV above the Fermi level. For SLy4, the value $V_0 = -1250$ MeV fm$^{-3}$ has been used (see Ref. [21]) based on the properties of SD bands in the mass $A \approx 150$ region, with a cut-off in the pairing space of 5 MeV both above *and* below the Fermi level. It has been shown [16,21,22] that such an optimized model nicely reproduces high spin properties of SD bands in the $A \approx 150$ and 190 mass regions. For the SkP interaction, we have determined a strength $V_0 = -900$ MeV fm$^{-3}$, with the same cut-off recipe as for SLy4. This value leads to similar pairing gaps as for the other two Skyrme forces.

## III. RESULTS

### A. Two-particle separation energies: semi-magic nuclei

In order to illustrate the ability of the present-day theoretical methods to describe the experimental two-particle separation energies, we performed several sets of calculations for ground-state configurations of semi-magic nuclei which are expected to be spherical.

Concentrating on the region of nuclei around the doubly-magic $^{208}$Pb, we present in Fig. 1 the two-neutron and two-proton separation energies in the $Z=82$ isotopes and $N=126$ isotones, respectively. These chains of semi-magic nuclei can be safely described by a spherical approximation. The self-consistent results obtained with the SkP and SLy4 interactions are compared with experimental values [23–25] and with the results of the macroscopic-microscopic Finite Range Droplet Model (FRDM) [26]. The SkM* results can be found in Ref. [15].

As seen in Fig. 1, the two-particle separation energies are reproduced with an overall accuracy of 1 to 2 MeV. In particular, the FRDM gives a very good description of the data; both the experimental two-neutron separation energies and the values extrapolated from systematic trends [23] are well reproduced. In the as-yet inaccessible region of heavy lead isotopes with $N\approx142$, the FRDM predicts a sudden appearance of deformation which gives rise to jumps in the $S_{2n}$ curve. The two-proton separation energies are described almost as well as the two-neutron separation energies; only the magnitude of the $Z=82$ shell effect is slightly underestimated by the FRDM.

Self-consistent models based on the Skyrme interaction do not perform so well in general, as discussed in Refs. [27,28]. However, around $^{208}$Pb the results obtained with the SkP force are fairly close to the data, except from slightly overestimated values of $S_{2n}$ just below the $N=126$ gap and slightly underestimated values of $S_{2p}$ just above the $Z=82$ gap. This force has an effective mass $m^*/m$ equal to one, similar to that used in macroscopic-microscopic methods, where it has been adjusted to specifically reproduce isotopic dependences of nuclear masses. The fact that the force SLy4 adopts a lower effective mass, $m^*/m=0.70$, is reflected in slightly overestimated shell effects at $N=126$ and $Z=82$. On the other hand, an effective mass of the order of 0.70 seems to be required by other microscopic arguments [29–31]. (The effective mass of SkM$^*$ is $m^*/m=0.79$, i.e., intermediate between the values for SLy4 and SkP.)

Before discussing results for spherical and deformed even-even nuclei with $110\leq N\leq 116$ and $78\leq Z\leq 82$, we note that the spherical self-consistent calculations reproduce very accurately the two-neutron separation energy in $^{194}$Pb and slightly overestimate the two-proton separation energy in $^{208}$Pb. The quality of data reproduction for $^{208}$Pb is comparable to that obtained within the FRDM which slightly underestimates the value of $S_{2p}$.

### B. Two-particle separation energies: first well

To analyze the ground-state two-particle separation energies in nuclei which are not semi-magic, one has to explicitly consider the deformation effects. Since the present study addresses questions related to *both* isotopic trends *and* deformation, we discuss results obtained with all the three forces, as they are focused on either one of these two particular aspects.

The calculated ground-state two-neutron separation energies in even Pt, Hg, and Pb isotopes with neutron numbers between $N=110$ and 116 are shown in Fig. 2. Figure 3 displays the calculated $S_{2n}$ values for the even-even U and Pu isotopes with $140\leq N\leq 146$. When confronting theoretical results with experiment we use the recent Schottky mass measurements at the GSI ESR [24,25,32] which generally confirm the systematic values of Ref. [23]. Based on these results, several conclusions can be drawn. For the SkM$^*$ force, the agreement with experiment is rather poor; in all cases, SkM$^*$ leads to an overestimation of the data by 0.5-1 MeV. This confirms earlier observations [15] that SkM$^*$ does not reproduce correctly the isotopic dependence of nuclear masses. On the other hand, the SkP interaction reproduces the data very well in all cases, as does the FRDM. The disagreement in $^{188}$Pt obtained in the FRDM results from a sudden change of deformation predicted by this model in this nucleus [26]; such an effect is not obtained with the Skyrme forces. The SLy4 force gives a good data reproduction in the Hg region while it slightly underestimates the $S_{2n}$ values in the U and Pu isotopes.

When it comes to the two-proton separation energies, the pattern obtained for various forces is different (see Fig. 4). The SkM* interaction does very well for the Hg isotopes, it slightly underestimates the data for the Pb isotopes, and fails rather badly for $Z=94$. The results obtained with SkP are of similar quality as those obtained with SLy4: the former gives a very good description of the data for the Hg isotopes, the latter gives an excellent agreement with the U data. In the Pb isotopes both overestimate the experimental $S_{2p}$ values by $\sim 1$ MeV. The overall quality of data reproduction by the FRDM is slightly better than for the SkP and SLy4 models, although a failure to reproduce the Pb chain is to be noted.

Up to now, no direct constraints on the surface energy has been introduced in the adjustments procedures of the Skyrme forces. The SkM* parametrization is the only one for which a deformation property has been included in the fit. Many properties of the forces have been adjusted to the global nuclear matter properties, such as volume and symmetry energies. These measures seem to be too crude when describing experimental data at the level of accuracy below 1 MeV. More important are probably positions of individual single-particle levels which crucially influence the deformations and deformation energies, and hence the ground-state separation energies. In spite of these qualifications, both SkP and SLy4 perform surprisingly well, and their very different effective masses do not seem to affect the quality of agreement with data. It is also clear that the SkM* interaction is probably not the optimal choice when describing isotopic variations of binding energies.

### C. Excitation energies of superdeformed minima

In several previous works based on macroscopic-microscopic methods and self-consistent approaches, excitation energies of SD minima have been predicted (see Ref. [33] for a review). However, since the excitation energy involves a difference between the binding energies of SD and ground-state minima, it can easily be obscured by a different quality of the theoretical description for such different states. Here, particular properties of Skyrme parametrizations that determine the deformability of a nucleus, such as the surface tension, may play a significant role. Another source of uncertainty concerns the corrections which should be added to the calculated energies to account for rotational symmetry breaking. For some interactions, for example SkM* and D1S [34], selected deformation properties (e.g., fission barriers) have been included in the global fit of force parameters, assuming no rotational corrections. In such cases, one assumes that all the corrections due to deformation have been effectively included. In other cases (see, e.g., Ref. [35]) predicted masses have been corrected for the rotational zero-point energy. Finally, for some interactions such as SkP and SLy4, only properties of spherical nuclei have been considered. In most cases, results of deformed calculations are not corrected for the rotational zero-point motion.

The excitation energies of the SD minima, $E_{SD}$, calculated in this work are shown in Figs. 5 (Pt, Hg, and Pb) and 6 (U and Pu). It is seen that (i) the predicted values rather strongly depend on the interaction and (ii) none of them does a particularly good job, the SkM* results being closest to the data. The disagreement is particularly striking for the actinides where the SkP and SLy4 forces overestimate experimental values of $E_{SD}$ by more than 2 MeV. On the other hand, considering the uncertainties discussed above, the excitation energy of a SD state is not a very useful characteristic of the model; small model variations can result in large changes in $E_{SD}$. For instance, the values of $E_{SD}$ predicted for the actinide nuclei in Ref. [19] with the same SkM* force as in this work are by ∼1MeV lower than our results. This difference can probably be attributed to a different treatment of the pairing channel. Namely, a seniority force within the HF+BCS method approximation was used in Ref. [19]. One should note that this latter choice is closer to the pairing treatment adopted in the fitting procedure of SkM*.

The accuracy of self-consistent methods for reproducing the absolute ground-state energies of heavy deformed nuclei is considerably less than for the relative energies [28]. For example, the values of binding energies $B_{GS}=-E_{GS}$ are in $^{238}$U underestimated by about 7.4, 6.6, and 10.8 MeV for SkP, SLy4, and SkM*, respectively. Some part of this discrepancy can be attributed to the numerical algorithms used in the present calculations, namely the finite-difference treatment of the kinetic energy. The resulting systematic error is expected to *increase* the deviation between experiment and theory by additional 3 to 4 MeV. These large errors suggest that the absolute energies should be used with caution when assessing merits of effective forces used in the self-consistent calculations. On the other hand, the relative energies (e.g., particle separation energies or deformation energies) are reproduced much better, and hence are more useful for assessing the quality of the effective interactions.

Figure 7 displays the calculated potential energy curves for $^{238}$U and $^{240}$Pu as functions of the total quadrupole moment $Q_{20}$. In both nuclei, the energies are shown relative to the ground-state energy $E_{GS}$. The axial barrier heights obtained for $^{240}$Pu ($^{238}$U) are 11.6 MeV (11.1 MeV) for SLy4, 10.5 MeV (9.7 MeV) for SkP, and 9.1 MeV (9.0 MeV) for SkM*. According to the analysis of Ref. [36], experimental inner barriers in $^{240}$Pu and $^{238}$U are ∼5.7 MeV and ∼5.6 MeV, respectively. For a meaningful comparison between experiment and theory, however, one should take into account the effect of triaxiality. For SLy4, the inclusion of nonaxial degrees of freedom reduces the inner barrier in $^{240}$Pu by 2.1 MeV; a slightly smaller effect was obtained for SkM* in Ref. [19]. Hence it can be concluded that all the Skyrme parametrizations employed in this work overestimate barrier heights in $^{240}$Pu and $^{238}$U by roughly 3 to 4 MeV. A similar conclusion has been reached in Ref. [37] with non-relativistic and relativistic calculations.

Considering the above uncertainties, it seems much safer to concentrate on the energy differences between SD minima, i.e., particle separation energies in the second well. For those quantities, involving energy differences between SD states only, one can hope that dynamical effects and treatment of pairing correlations would play a less important role.

### D. Two-particle separation energies: second well

When studying the separation energies in the SD configurations, one expects that theoretical predictions should be robust as they depend on general properties of effective interactions. This fact may have its roots in specific symmetry properties of SD states [38,39] and has been noticed in several theoretical studies of SD high-spin bands using the HF method [40–42]. Energy relations between the ground states and SD minima in three adjacent even-even nuclei $^{192}$Hg, $^{194}$Hg, and $^{194}$Pb are schematically presented in Fig. 8. It shows three potential energy curves in these nuclei, approximately shifted in energy according to the ex-perimental two-particle separation energies. The absolute binding energies of ground states and SD minima define six points in the absolute energy scale, hence seven interesting energy differences, which are indicated in Fig. 8 by straight dotted lines. These are three excitation energies $E_{\rm SD}$ of SD minima in the three considered nuclei as well as the two-neutron and two-proton separation energies in normal and SD minima. Given the uncertainty in the excitation energy $E_{\rm SD}$ in $^{192}$Hg, the mass difference for the SD states in $^{194}$Pb and $^{194}$Hg is also an interesting quantity.

Values of the energy differences of Fig. 8 are shown in Fig. 9. Experimental data are presented in panel (a), while panels (b), (c), and (d) show deviations between the theoretical and experimental results. One should note that the experimental separation energies in Fig. 9 are taken from Ref. [24,25] and have error bars of the order of 0.1 MeV. An uncertain piece of experimental data is the excitation energy of the SD minimum in $^{192}$Hg. Unfortunately, in spite of several experimental efforts, the direct link between SD bands and the known yrast line in $^{192}$Hg has not yet been found. The number quoted in Fig. 9 is based on estimates from on-going analysis [3]. The only firmly established quantity is the binding energy difference between the SD minima in the $A$=194 isobars of Hg and Pb, i.e., the difference between two-proton and two-neutron separation energies $S_{2p}$−$S_{2n}$.

The general observation following results shown in Fig. 9 is that the self-consistent calculations reproduce the data to within 1–2 MeV. This seems to be true for both ground states and SD states. As far as the SD minima are concerned, both SLy4 and SkP give an excellent agreement with the experimental value of $S_{2p}$−$S_{2n}$, Table I.

In spite of this agreement, the analysis of results shown in Fig. 9 suggests that there is still room for improvement. For the forces SkP and SLy4, the pattern of desired modifications is quite clear. As illustrated in Fig. 9 by thick arrows and numbers in ovals, a significant improvement of results would have been obtained if the theoretical energies of the SD state in $^{192}$Hg were raised (∼0.8 MeV for SkP and ∼0.5 MeV for SLy4) and the ground-state energies of $^{194}$Pb were raised by a similar amount (∼0.9 MeV for SkP and ∼1.3 MeV for SLy4). Changes of that order would bring the agreement with the experimental data to the level of 0.5 MeV. Due to the uncertain experimental value of $E_{\rm SD}$ in $^{192}$Hg we may speculate that a value lower by about 0.6 MeV would result in a very consistent picture for the SkP and SLy4 forces. Namely, in such a situation the only significant remaining discrepancy would be the ground-state energy of a semimagic spherical system $^{194}$Pb.

For SkM*, the pattern of changes is different and the magnitude of deviations is larger. In particular, it seems that in $^{194}$Pb both the ground-state and the SD state energies should be shifted. This suggests that the good reproduction of the excitation energy of SD minimum in $^{194}$Pb by this model (see Fig. 5) is *fortuitous*.

For the shape isomers in actinide nuclei, energies have been reported for $^{236}$U, $^{238}$U, but only an approximate energy is known for $^{240}$Pu. One can construct and analyze similar binding energy differences as for the $^{194}$Hg, $^{192}$Hg, and $^{194}$Pb nuclei. In particular, one can deduce three excitatation energies $E_{\rm SD}$ and the two-proton and two-neutron separation energies in shape-isomeric states. Again, both SLy4 and, in particular, SkP give a good agreement with the experimental values of $S_{2p}$ for the SD minimum in $^{240}$Pu and $S_{2n}$ for the SD minimum in $^{238}$U, Table I. The force SkM* gives a very poor description of the SD separation energies for both neutrons and protons.

## IV. SUMMARY

The self-consistent Skyrme-HFB method has been applied to investigate binding energy relations in the first and second wells of nuclei in the $A\sim190$ and $A\sim240$ mass regions. For the two-nucleon separation energies *within* the first and *within* the second minimum, the Skyrme interactions that have been optimized for isotopic trends, SLy4 and SkP, give a good agreement with experimental data, irrespective of their very different effective masses. For the ground-state separation energies, the level of data reproduction by these forces is similar to that obtained with the macroscopic-microscopic method. For the limited number of binding energy differences in the SD minima known experimentally, the agreement between theory and experiment is good for both SLy4 and SkP. The SkM* parametrization has been confirmed to have wrong isospin behavior and should not be considered when making binding-energy extrapolations.

The good agreement obtained for particle separation energies does not hold for *relative* energy differences between the first and second well. Namely, neither SLy4 nor SkP have been able to reproduce the excitation energy of the SD minimum, although their predictions are very close with each other. One has to bear in mind, however, that the values of $E_{\rm SD}$ are sensitive to model uncertainties such as treatment of pairing or dynamical zero-point correlations [37,43].

The SkM* parametrization fails in reproducing isotopic trends. It does slightly better for fission barriers and $E_{\rm SD}$. Clearly, the inclusion of a deformation effects during the fitting procedure of force parameters should have an effect on the predictive power of the force at large deformations. Another necessary improvement would be a simultaneous optimization of the mean-field and the pairing field. Such a philosophy has been adopted for the D1S and SkP interactions, which are simultaneously used in both the particle-hole and pairing channels. Also, the dynamical zero-point corrections should be consistently considered when optimizing force parameters. The latter corrections are expected to be of particular importance when comparing energies of minima with very distinct intrinsic structures and deformations.

This work has been motivated by the recent experimental data on absolute energies of SD bands in the $A\sim190$ mass region. Although very scarce, the experimental data on separation and excitation energies have already shed a new light on basic properties of effective interactions such as the isospin dependence and the response of the system to shape deformations.

## ACKNOWLEDGMENTS

We would like to thank H. Wollnik and T. Radon for providing results of their mass measurements prior to publication. This research was supported in part by the U.S. Department of Energy under Contract Nos. DE-FG02-96ER40963 (University of Tennessee), DE-FG05-87ER40361 (Joint Institute for Heavy Ion Research), DE-AC05-96OR22464 with Lockheed Martin Energy Research Corp. (Oak Ridge National Laboratory), W-31-109-ENG-38 (Argonne National Laboratory), by the Belgian Ministry for Science Policy (SSTC) under Contract No. 93/98-166, by the NATO Grant CRG 970196, and by the Polish Committee for Scientific Research.

TABLES

TABLE I. Particle separation energies (in MeV) in superdeformed minima calculated with three Skyrme interactions and compared with available experimental data.

|      | $S_{2p}(^{194}\text{Pb})$—$S_{2n}(^{194}\text{Hg})$ | $S_{2p}(^{240}\text{Pu})$ | $S_{2n}(^{238}\text{U})$ |
|------|---------------------------------------------------|---------------------------|--------------------------|
| SkM* | 10.3 | 10.0  | 12.4  |
| SLy4 | 8.3  | 12.1  | 10.5  |
| SkP  | 8.5  | 11.4  | 11.1  |
| EXP  | 8.20 | 11.52 | 11.47 |

FIGURES

FIG. 1. Two-neutron ground-state separation energies in the chain of lead isotopes (top) and two-proton separation energies in the chain of $N=126$ isotones (bottom). Solid and dashed lines show the results obtained with the SLy4 and SkP forces, respectively, using the spherical HFB approach. The FRDM results of Ref. [26] (dash-dotted line) are also given.

FIG. 2. Two-neutron ground-state separation energies in the even-even Pt, Hg, and Pb isotopes with the neutron numbers between $N=110$ and 116 obtained in the deformed HFB+LN model. Solid, dashed, and dotted lines show the results obtained with the SLy4, SkP, and SkM* forces, respectively. They are compared to the results of the FRDM approach [26] (dash-dotted line) and to experimental data [23–25].

FIG. 3. Same as in Fig. 2 but for the even-even U and Pu isotopes with $140 \leq N \leq 146$.

FIG. 4. Same as in Fig. 2 but for the two-proton ground state separation energies in the even-even Hg, Pb, and Pu isotopes.

FIG. 5. Excitation energies of the SD minima with respect to the ground-states of the even-even Pt, Hg, and Pb isotopes with the neutron numbers between $N=110$ and 116. Solid, dashed, and dotted lines show the results obtained with the SLy4, SkP, and SkM* forces, respectively. Experimental data are taken from Ref. [2] ($^{194}$Pb; 4.64 MeV) and [1] ($^{194}$Hg; 6.01 MeV). The tentative points for $^{192}$Hg (5.4 MeV) [3] and $^{192}$Pb (3.9 MeV) [4] are also shown.

FIG. 6. Same as in Fig. 5 but for the even-even U and Pu with $140 \leq N \leq 146$. Experimental data are taken from Ref. [44] ($^{236}$U; 2.75 MeV), [45,46] ($^{238}$U; 2.56 MeV), and [47,46] ($^{240}$Pu; 2.8 MeV). Only an approximate value is known for $^{240}$Pu.

FIG. 7. Potential energies as functions of the quadrupole moment for $^{238}$U (top) and $^{240}$Pu (bottom) calculated with the SkP (dashed line), SLy4 (solid line), and SkM* (dotted line) Skyrme forces. The values are normalized to zero at the ground-state energies $E_{\rm GS}$.

FIG. 8. Schematic representation of potential energy curves in three adjacent even-even nuclei $^{192}$Hg, $^{194}$Hg, and $^{194}$Pb. Dotted straight lines indicate the relative binding energy relations in and between these nuclei. These are the two-particle ground-state separation energies, the two-particle SD separation energies, and the SD excitation energies. The same straight lines are reproduced in Fig. 9 below where the calculated and experimental values for each of these energies are given.

FIG. 9. Values of the relative energies shown in Fig. 8. Experimental values (a) are compared with the results of the self-consistent HFB+LN calculations with SLy4 (b), SkP (c), and SkM* (d) effective interactions. The numbers in panels (b)–(d) give the differences between theory and experiment. The value of $E_{\mathrm{SD}}$ for $^{192}$Hg in panel (a) has not been measured, but is the current estimate from on-going analysis [3]. For clarity all values have been rounded to 0.1 MeV. See text for the meaning of thick arrows and numbers in ovals.

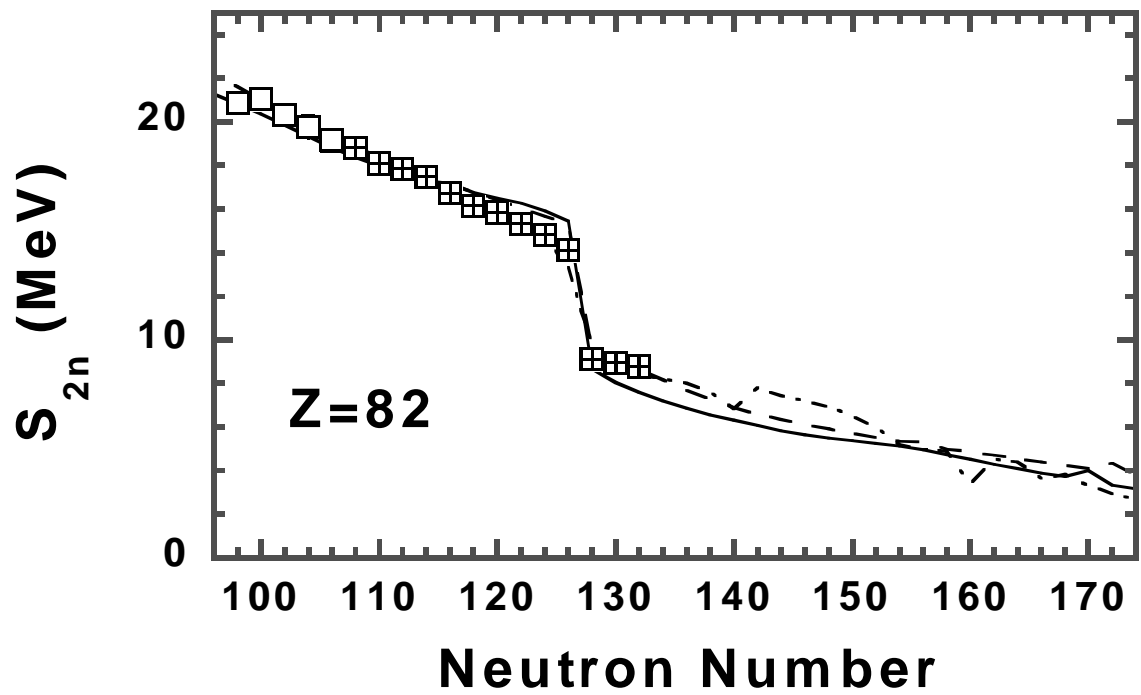
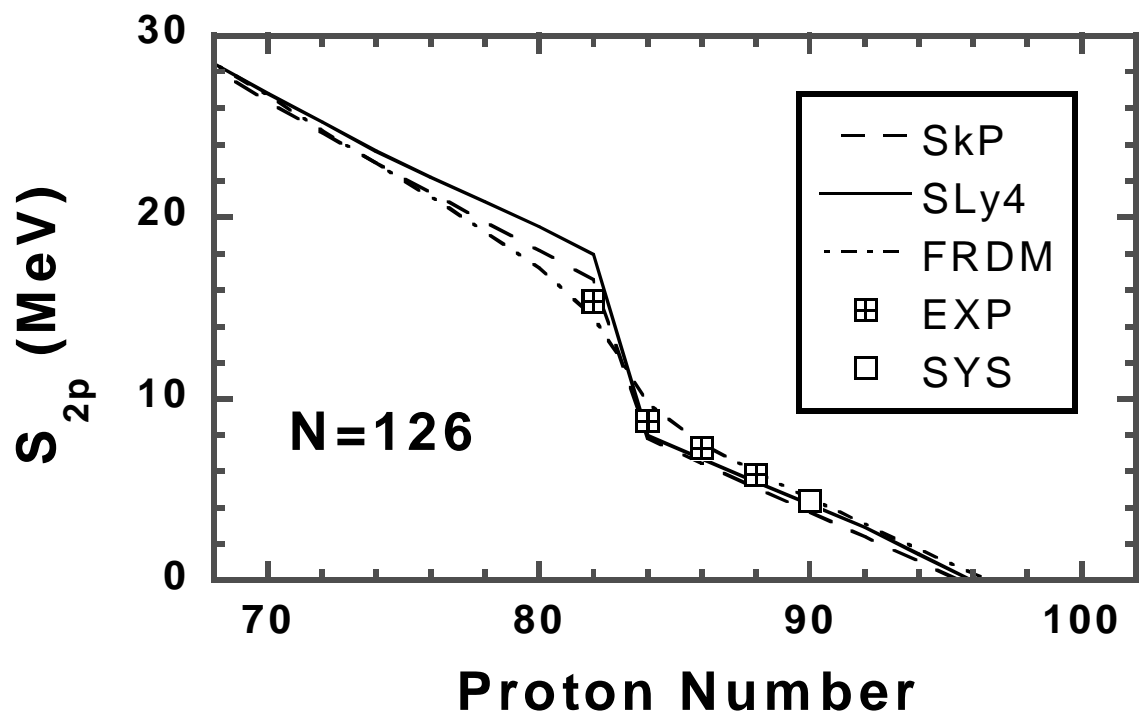

**Figure 1**

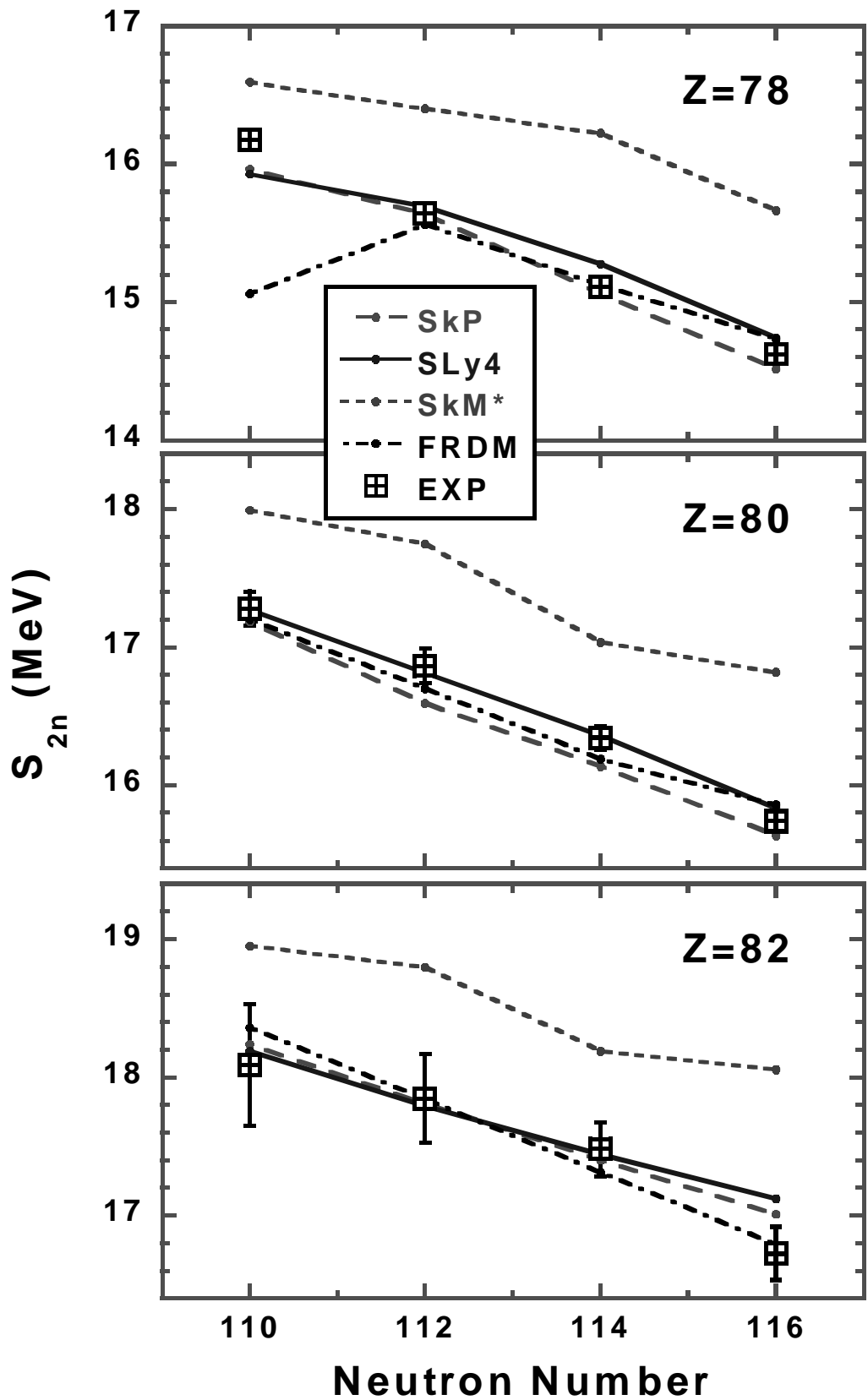

Figure 2

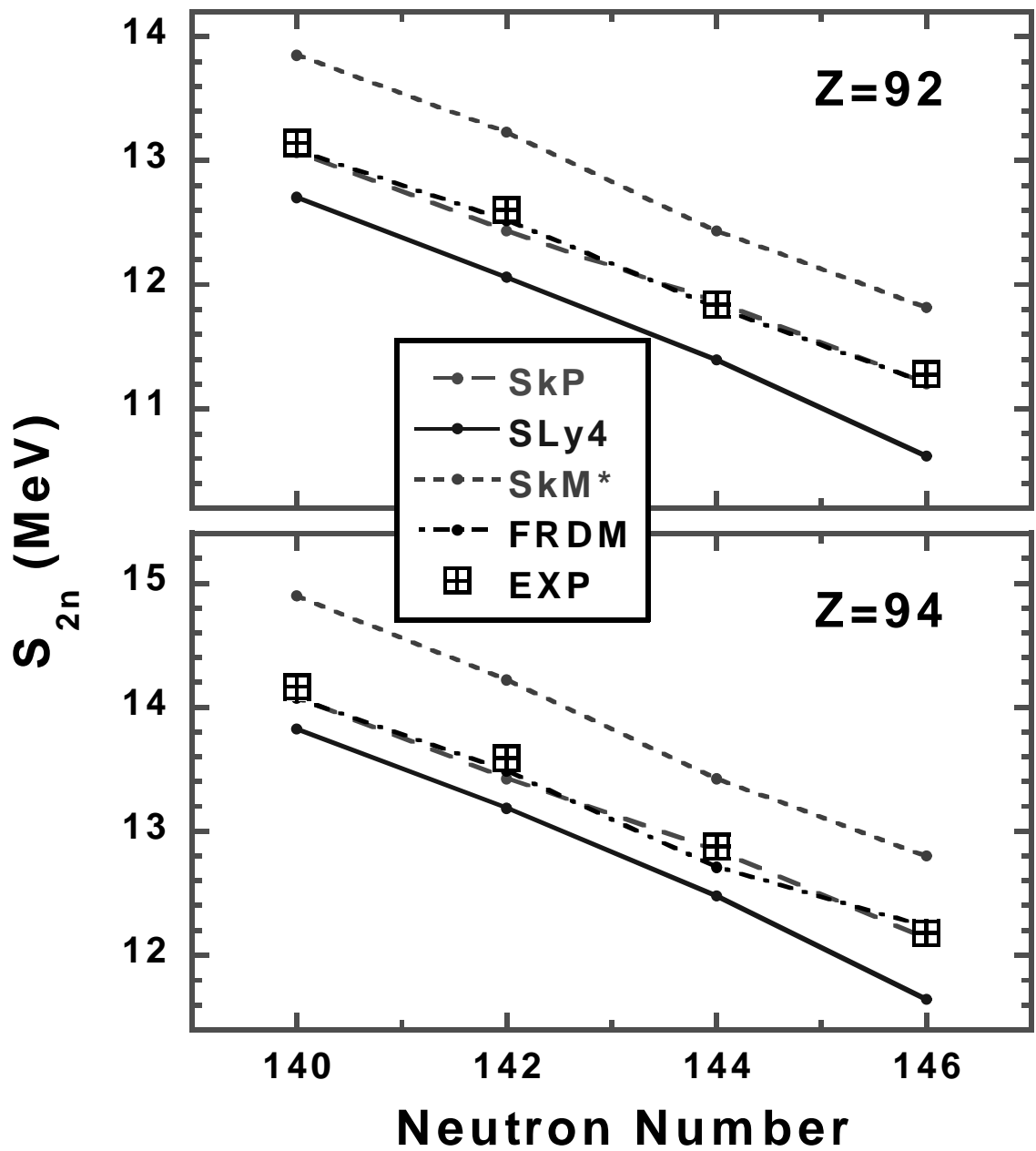

Figure 3

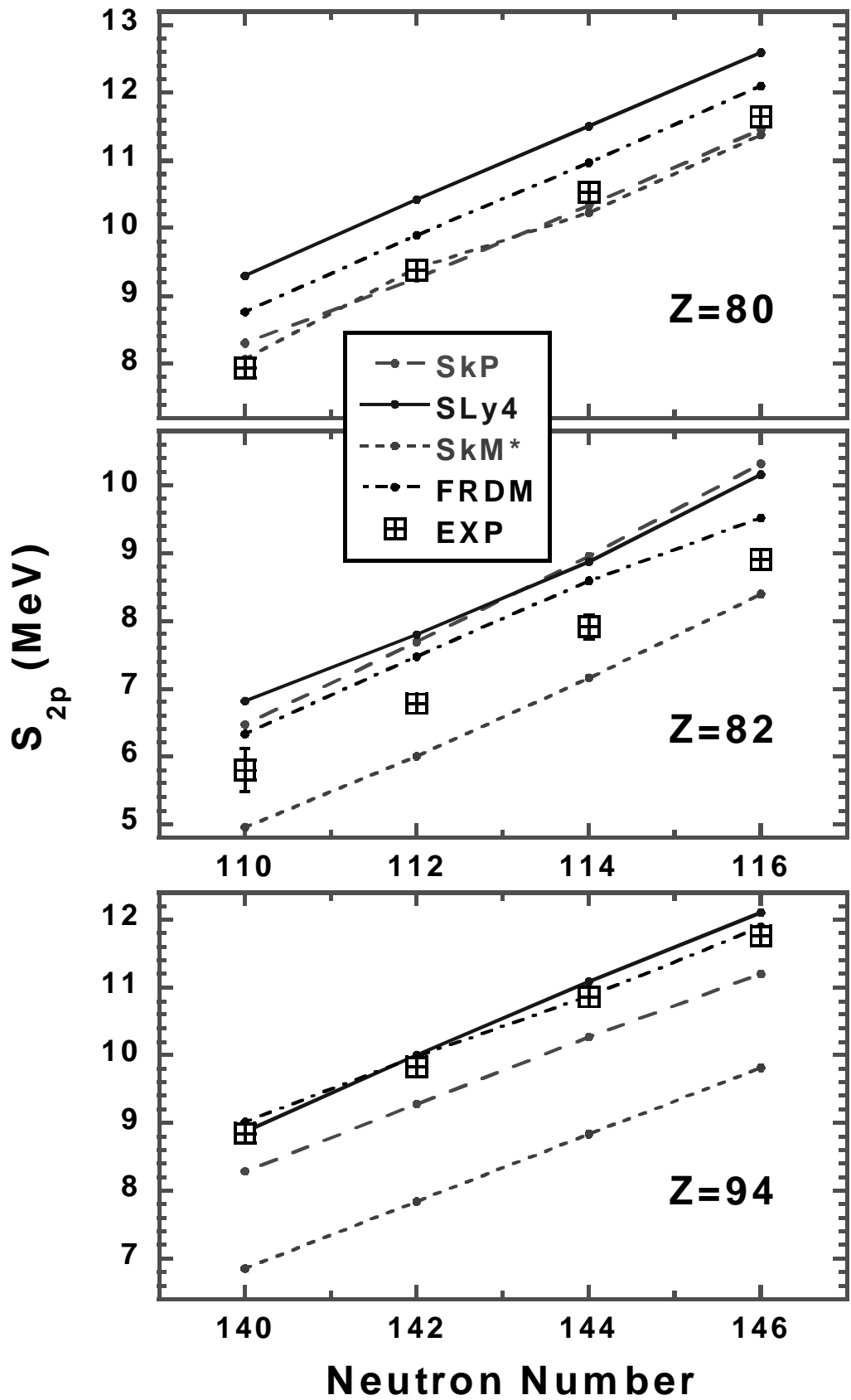

Figure 4

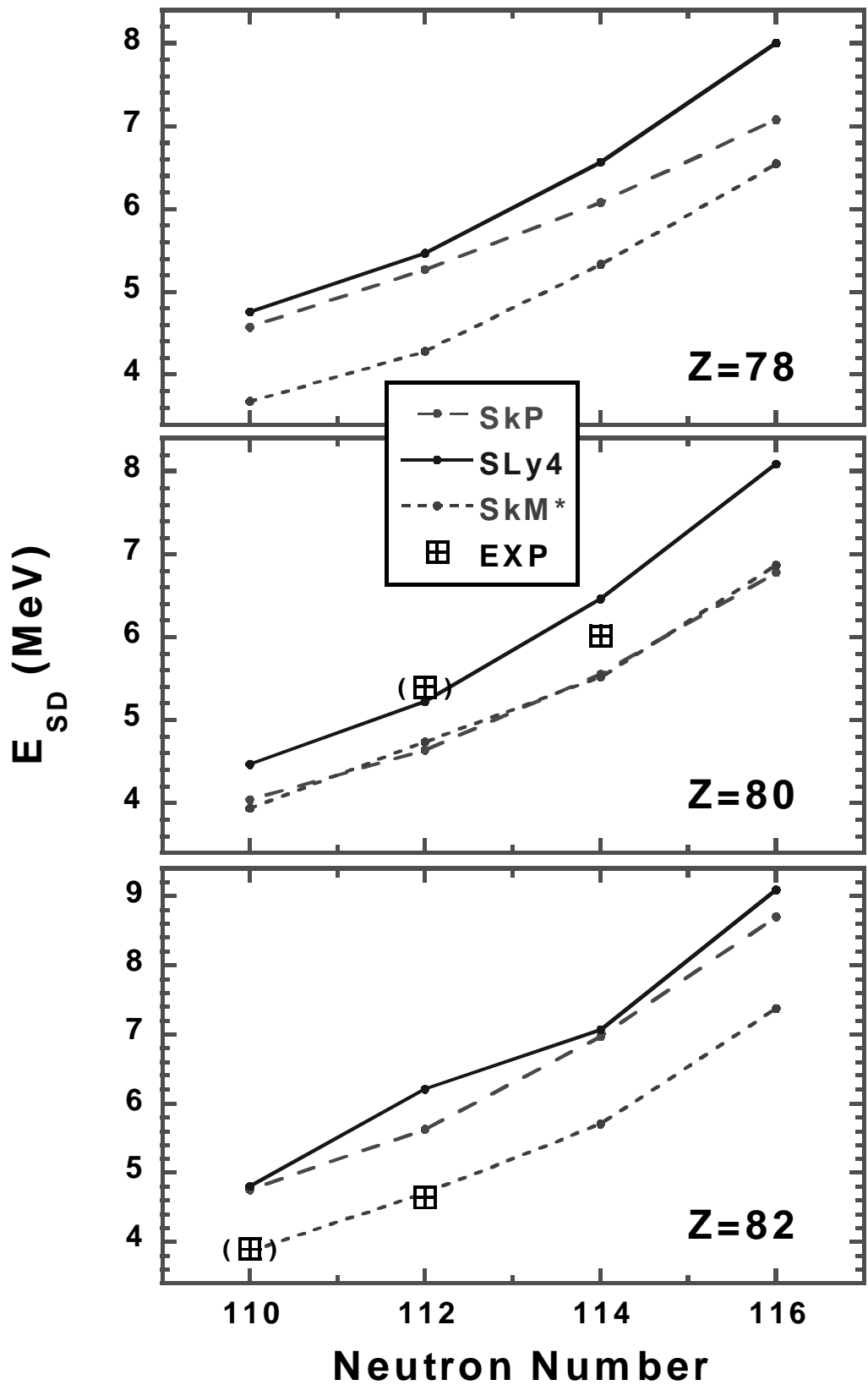

**Figure 5**

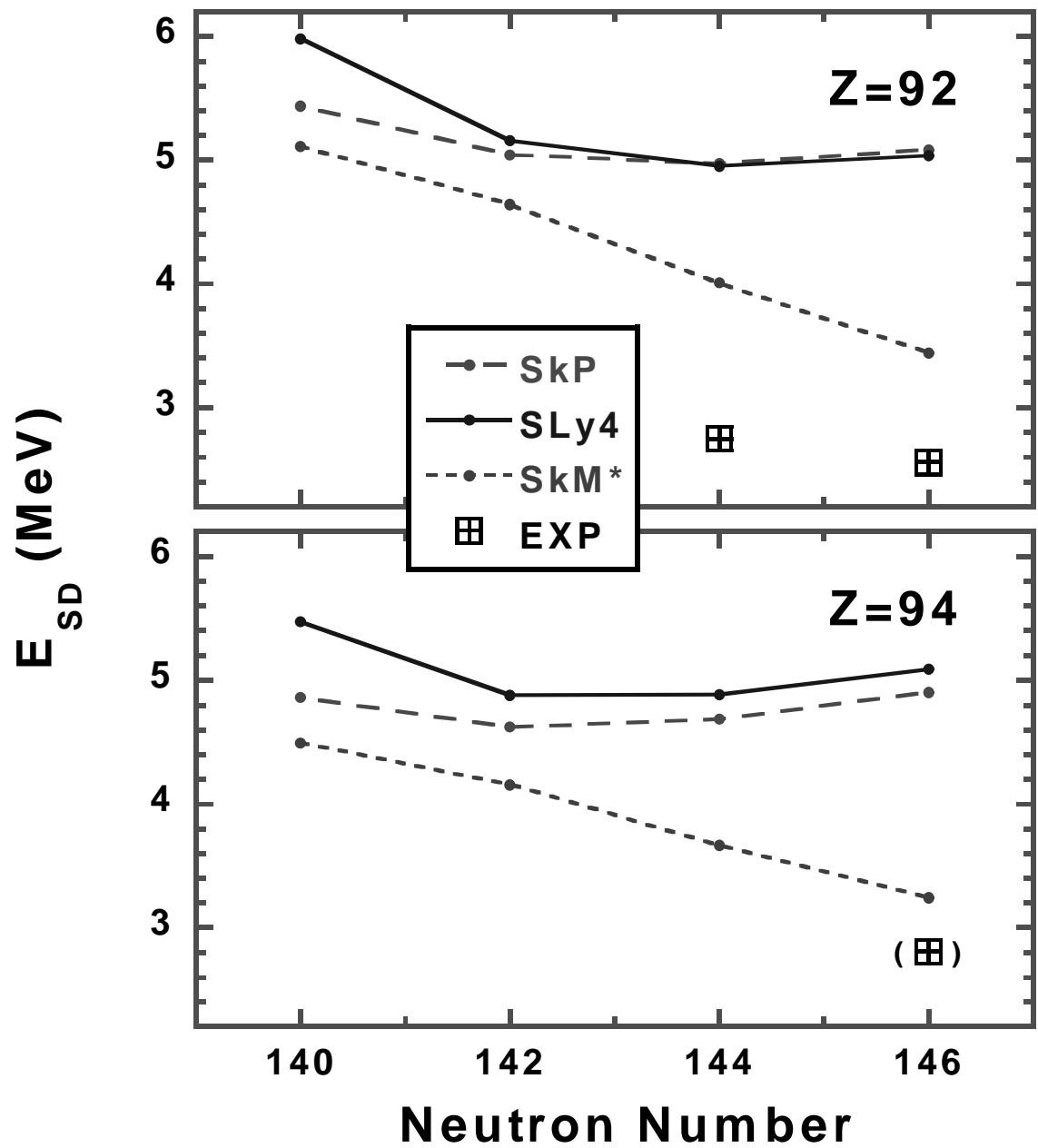

**Figure 6**

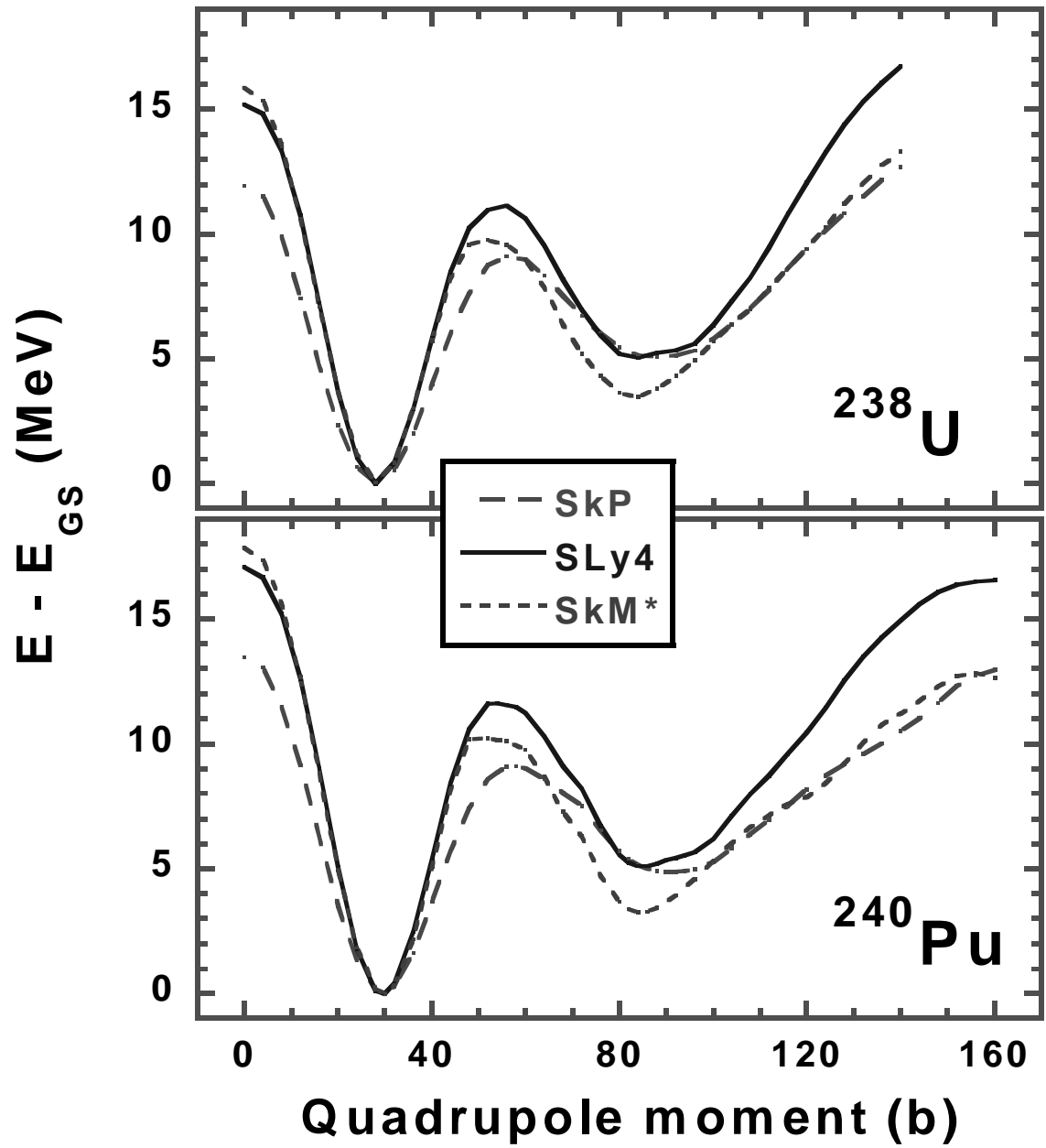

**Figure 7**

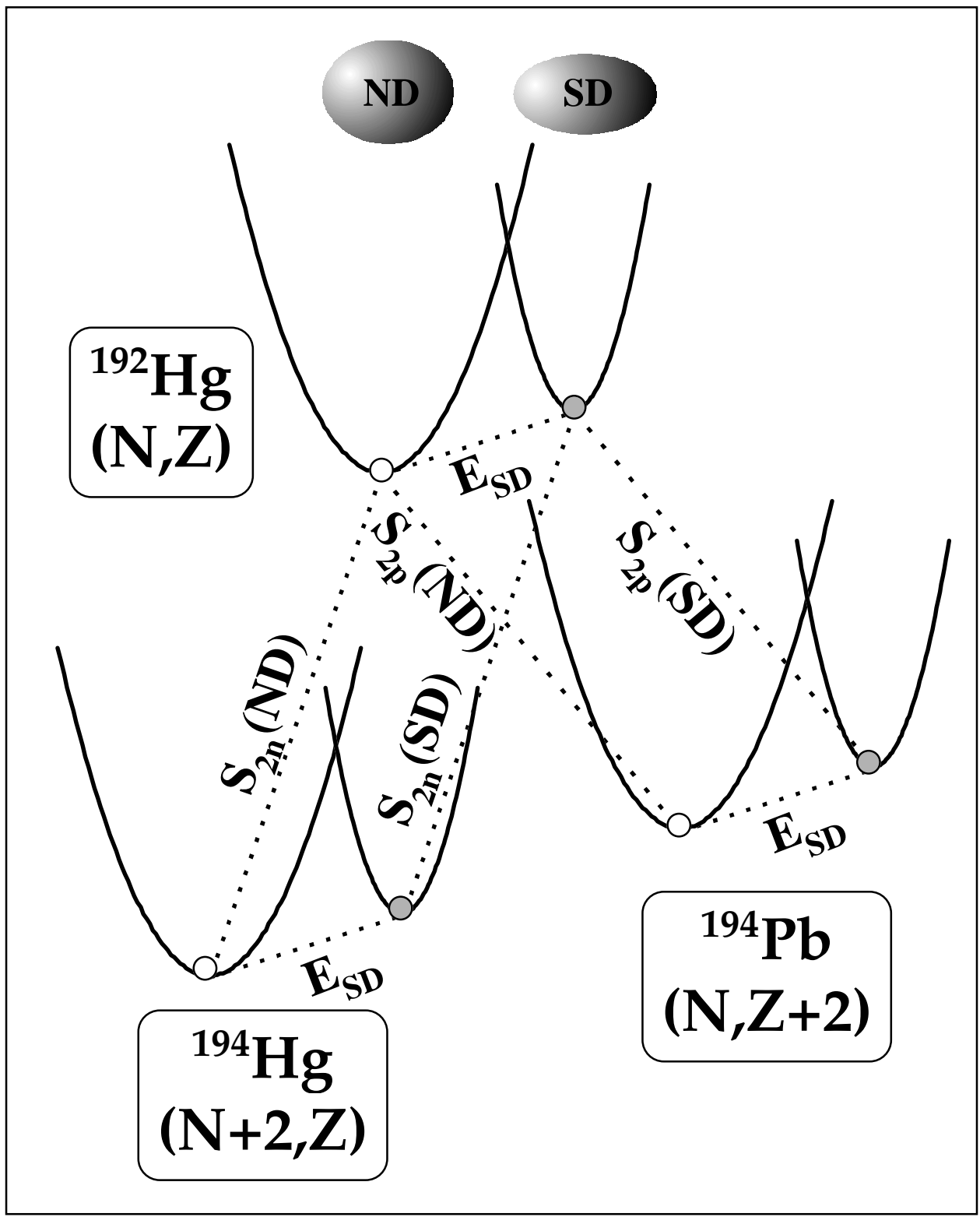

Figure 8

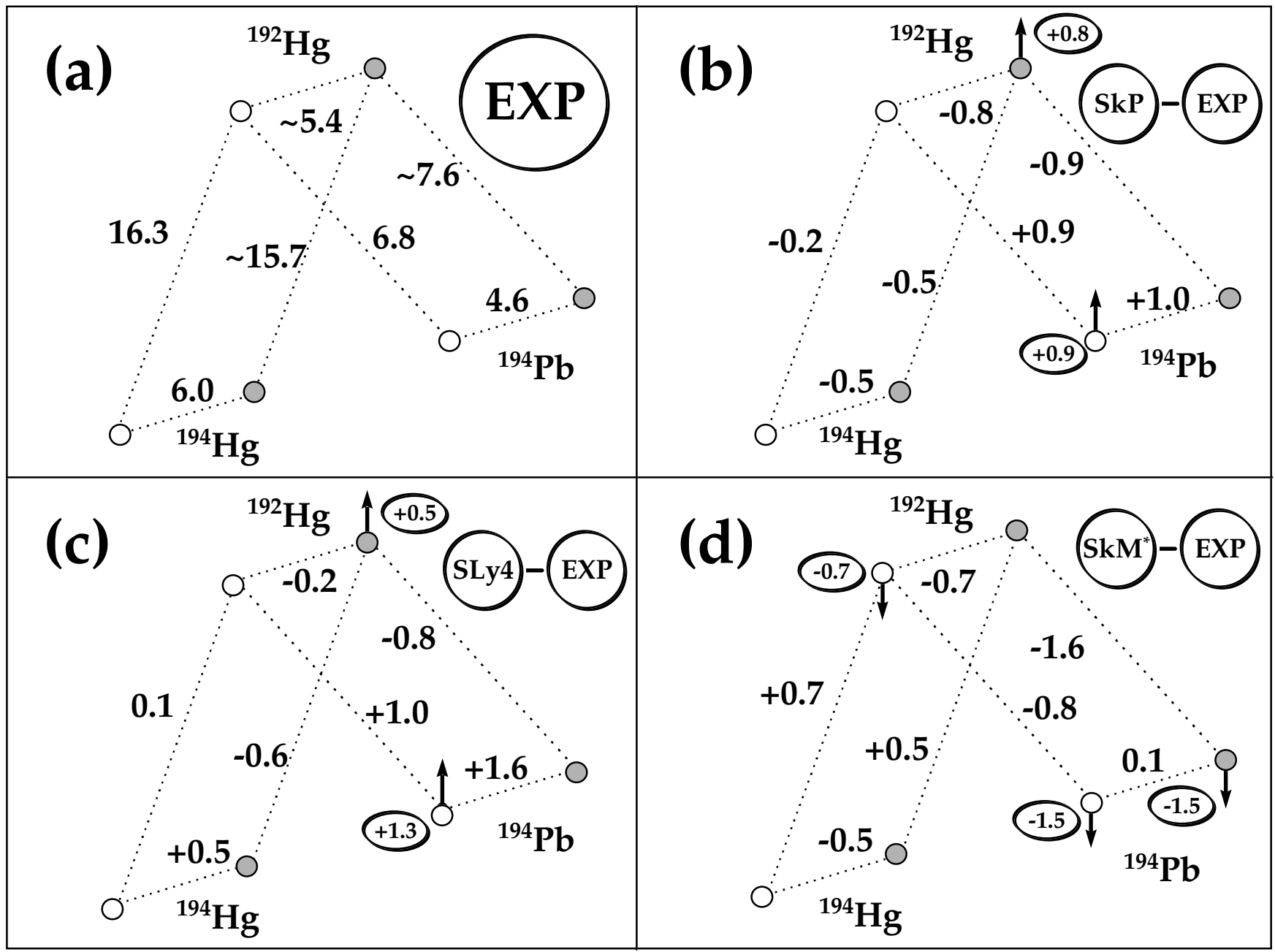

Figure 9